\documentclass[10pt, dvips]{article}
\usepackage{epsfig}
\begin{document}

\title{\bf A hierarchy of cosmic compact objects - without black holes}

\author {Johan Hansson \\
 Department of Physics, Lule{\aa} University of Technology
 \\ SE-971 87 Lule\aa, Sweden}

\date{}

\maketitle

{\bf We make the case for the existence of a, hitherto unknown and
unobserved, hierarchy of ever more compact cosmic objects in the
universe. This hypothesis is based on i) the assumption of
"elementary" particle sub-constituents on several levels below the
presently known, inspired by Glashow's "blooming desert"
\cite{Glashow}, ii) the existence of nearly scale-invariant
density fluctuations in the early universe, e.g. as predicted by
inflationary models \cite{Inflation,scaleinv}, iii) our own
previous theoretical work showing that a class of objects
considerably more compact than previously thought possible in
astrophysics can exist \cite{HanssonSandin}. We also give several
independent arguments strongly pointing towards the non-existence
of black holes. Some brief suggestions on observational signals
due to the hierarchy, both in collected astronomical data and in
possible future observations, concludes the paper.}


In our point of view it is extremely naive to assume a huge
unpopulated "desert" in the immense region separating the scale of
the standard model in particle physics ($\sim 10^{-18}$ m),
presently our most fundamental \textit{tested} description of
nature \cite{PDG}, and the grand unified (GUT) \cite{GUT1,GUT2}
($\sim 10^{-31}$ m) and/or superstring
\cite{superstring,Polchinski} ($\sim 10^{-35}$ m) scale. The
difference in size between a superstring and an atom is roughly
the same as the difference between an atom and the solar system,
and the question on the existence of atoms was just being resolved
only 100 years ago. Also, historically no extrapolations of then
reigning theories, as far into the unknown as the one leading to
the hypothetical "desert" have been successful. Instead,
successive layers of substructure have been found as shorter and
shorter distances have been probed experimentally. Inspired by
Sheldon Glashow \cite{Glashow} who coined the term "desert" only
to then renounce it, we envisage a fertile "blooming desert"
landscape teeming with substructure at all scales, awaiting to be
discovered in the future. "Today we can't exclude the possibility
that micro-unicorns might be thriving at a length scale of
$10^{-18}$ cm." \cite{Desperately}

Also, many models of cosmic inflation in the very early universe
give "fractal-like" \cite{Mandelbrot} (scale-invariant) density
fluctuations \cite{Inflation,scaleinv}. It is known that the
density fluctuations in the early universe act like "seeds" for
gravitational structure formation, once the fluctuations are
there, gravity does the rest. The difference between over-density
and under-density regions will automatically increase due to the
expansion of the universe, contracting gravitationally bound
systems and diluting gravitationally unbound (expanding) systems.
Gravitationally overdense regions act like "mini-universes" of
positive curvature, expanding to a maximum size and then
recollapsing. The larger the density contrast and the smaller the
size, the shorter the "mini-universe" lifetime and the smaller its
total mass.

The two preceding paragraphs implicate that gravitationally
induced structures should exist on all length scales, at least
those being stable. We have previously shown \cite{HanssonSandin}
that "preon stars" can exist, and are stable\footnote{Preons are
theoretically suggested sub-constituents of quarks and leptons
\cite{dSouza,Trinity}.}. There is nothing magical about neutrons
making them the last in line as constituents of cosmic compact
objects, as previously believed. But then there can be nothing
magical about preons either. If the "desert" of particle physics
in reality is populated by particles and sub-structure on many
scales, it must be possible for even more compact objects than
preon stars to exist, as stability will be assured in some corner
of parameter space. (The details will only alter the
\textit{abundance} of those objects, i.e. their number density in
a given volume of the universe.)

If we plot the known observed classes of gravitational structure,
a representative sample of which is shown in Fig.1, we see that
they always stay well away from the region of black
holes\footnote{Already in 1969, G. de Vaucouleurs observed a
universal density-radius relation for gravitational structures,
$log(\rho) = -21.7 -17(log(R) -21.7)$ (in cgs-units). "This leads
one to view the Hubble parameter as a stochastic variable, subject
in the hierarchical scheme to effects of local density
fluctuations on all scales." \cite{deV}}. Our qualitative
prediction for the hierarchy of cosmic compact objects is shown in
Fig.2. As can be seen, it does not include black holes.

\begin{figure}[h]
\begin{center}
\psfig{file=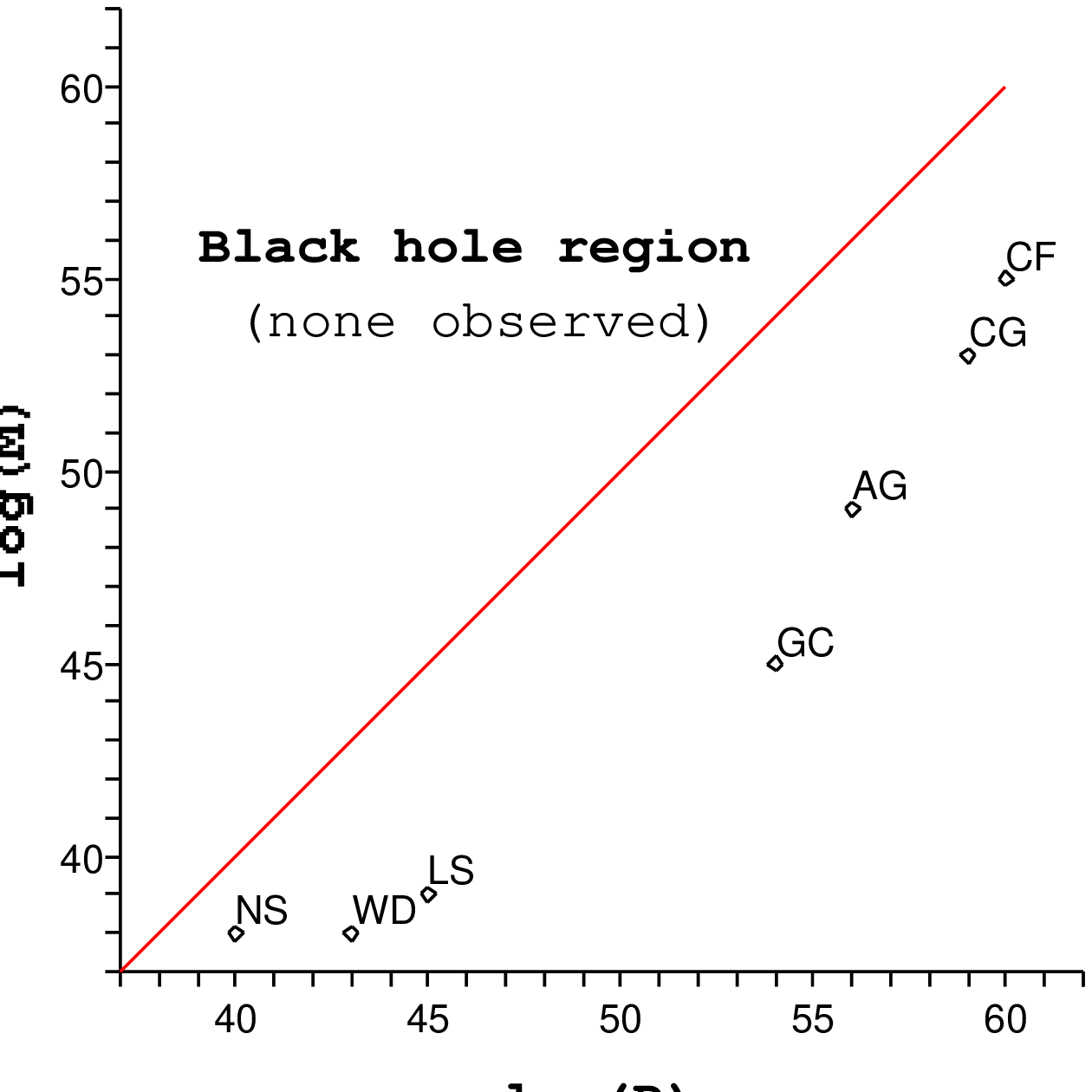}
\end{center}
 \caption{"The Universe on a line" - including known, representative
gravitational structures in the universe. NS - neutron star, WD -
white dwarf, LS - large star, GC - globular cluster of stars, AG -
average galaxy, CG - cluster of galaxies, CF - cosmic
filament/supercluster of galaxies. Gravitationally bound objects
larger than white dwarfs are included for completeness, they are
not compact, i.e. they are not held up exclusively by degeneracy
pressure from their quantum constituents. Masses and sizes are
expressed in dimensionless units $M = m/m_{Planck}$ and $R =
r/r_{Planck}$, where $m$ and $r$ are characteristic masses and
radii of the objects, while $m_{Planck} = \sqrt{\hbar c / G}$ and
$r_{Planck} = \sqrt{\hbar G / c^3}$ are the fundamental mass and
distance expected from a quantum theory of gravity.}
\end{figure}

There are, in fact, several independent hints, which when taken
together strongly point towards the conclusion that black holes
can never actually form. (See also \cite{Chapline}.) In classical
general relativity there is the "cosmic censorship" conjecture
\cite{WeakCosmicCensorship,StrongCosmicCensorship}, that
singularities are always shrouded from the outside world by event
horizons. We propose a much stronger cosmic censorship conjecture:
when fundamental microscopic physics is taken into proper account
it prevents the very formation and existence of black holes. Some
of the clues supporting our conjecture are

1) Traditionally \cite{Hawking}, the existence of black holes
inevitably causes a loss of quantum coherence, and hence a
breakdown of the basic principles of quantum mechanics,
essentially due to Hawking radiation \cite{HawkingRad}. One can
just as easily turn the argument around: As quantum mechanics is a
much more fundamental description of nature than general
relativity (which is merely a classical theory) the very same
chain of arguments, run backwards, implies that black holes can
never exist or form as it would violate quantum mechanics.

2) In a semi-classical, newtonian quantum gravity we can use the
mathematical identity (if $e^2/4 \pi \epsilon_0 \rightarrow GmM$)
of the Coulomb field and the newtonian gravitational field (mass =
gravitational "charge") to calculate the smallest allowed radius
of a quantum gravitational Bohr-model, for the limiting case that
all the mass-energy, $mc^2$, of an infalling test-particle becomes
binding energy, i.e. is radiated away in gravitons during infall,
making it impossible for the object to accrete more mass-energy.
It turns out that the gravitational "Bohr-radius" of any object is
of the order of the Schwarzschild radius in this case (for any $M$
as quantum theory is universal), hinting that black holes will not
form in the "old" quantum theory of Bohr, especially as all other
radiative mechanisms but gravity here have been neglected. (Seen
in the language of gravitational field lines, the gravitational
radiation effect will be larger the denser the field lines, as the
gradient then increases, but as the area of a spherical shell
increases in step with how the field lines decrease in $r$, the
effect is independent of the size of the presumptive black hole.
So even a very large black hole with an arbitrarily small actual
density and horizon curvature is prohibited.)

3) In analogy with the position of the electron in a Hydrogen
atom, Fig.3, the position of the event horizon for a black hole
will become fuzzy when more exact quantum mechanical effects are
taken into consideration. The Bohr radius for the Hydrogen atom,
and the Schwarzschild radius for a black hole, both dissolve when
subject to a more fundamental quantum mechanical treatment. The
"one-way membrane" of classical black holes gets penetrated, and
on a microscopic scale, dissolves completely. Seen another way,
particles can always quantum mechanically "tunnel" back through
the event horizon. In a like manner, the purely geometrical,
classical picture of point- (Schwarzschild) and ring-singularities
(Kerr) must dissolve when quantum mechanics is taken into account,
as simultaneously precise location and motion (zero!) is
forbidden. But as black holes are \textit{defined} by the very
presence of an event horizon and/or a singularity, we see that the
black hole itself cannot exist.

4) Quantum field theory and string theory arguments fare no better
as they presuppose the existence of a classical black hole
spacetime geometry, in which the fields/strings propagate, being
only small perturbations to the background geometry. What one
would need, but which so far is absent, is a derivation of black
hole-like solutions from first principles in a non-perturbative
theory of quantum gravity. According to our conjecture, such
solutions will never be found. (Unless quantum mechanics itself is
drastically altered.)

5) It is quite possible that radiation, both gravitational and
from particle physics (known and undiscovered), will always "bleed
off" sufficient mass-energy to keep any region of size $R$ below
the mass necessary even for the onset of classical collapse to a
black hole, $M < Rc^2/2G$ \footnote{This is strictly only valid
for a static, spherically symmetric mass distribution. The
relation serves just the purpose to illustrate our point. The
exact criterion for dynamical gravitational collapse in general
relativity is non-trivial, and unsolved in the general case.}. It
is known that the total energy emitted in e.g. the formation of a
neutron star in a supernova explosion is almost exclusively
carried away by neutrinos. There may well be "neutrino-like"
sub-species on the undiscovered "elementary" particle levels,
playing the same role in collapse. Such mechanisms could ensure
that $M < Rc^2/2G$ on all scales. (This could for example make it
possible for one, or several, "light" preon star(s) to form from a
very massive normal star, possibly population III
\cite{HanssonSandin}.) As an analogy, consider boiling water: if
we pump in a very large amount of energy, the water will not
attain a very large temperature, instead it will boil away in a
finite time. If in the analogy we replace boiling $\rightarrow$
radiating , water $\rightarrow$ mass-energy, temperature
$\rightarrow$ gravitational redshift, we get the gravitational
case. For a different viewpoint also reaching this conclusion, see
\cite{Chapline}.

6) Even if we presuppose the existence of a classical black hole,
infalling quantum mechanical "particles" (both matter and
radiation) behave like waves when unobserved. For particles
asymptotically falling in from infinity, the quantum mechanical
description will initially be an infinitely long harmonic wave.
Even though this will be altered when approaching the black hole,
part of their probability amplitude will always be outside the
classically defined event horizon, i.e. there will always be a
finite probability that the particle has \textit{not} fallen
through the horizon. In fact, the same applies for all the
particles that supposedly built up the black hole. According to
quantum mechanics, even nature itself can never tell if the
criterion $M = Rc^2/2G$ has been fulfilled, and thus does not know
if it is "supposed" to form a horizon and start collapsing the
interior. Because quantum mechanical "particles" are not just
simply microscopic pebbles, quantum mechanical black holes can
never form.

7) It may even be that gravity is only a macroscopically induced,
non-fundamental interaction, as proposed by Sakharov
\cite{Sakharov}. In that case, black holes are ruled out as
gravity is absent on microscopic scales, and the classical event
horizon and singularity must be defined with infinite, microscopic
precision.

\begin{figure}[h]
\begin{center}
\psfig{file=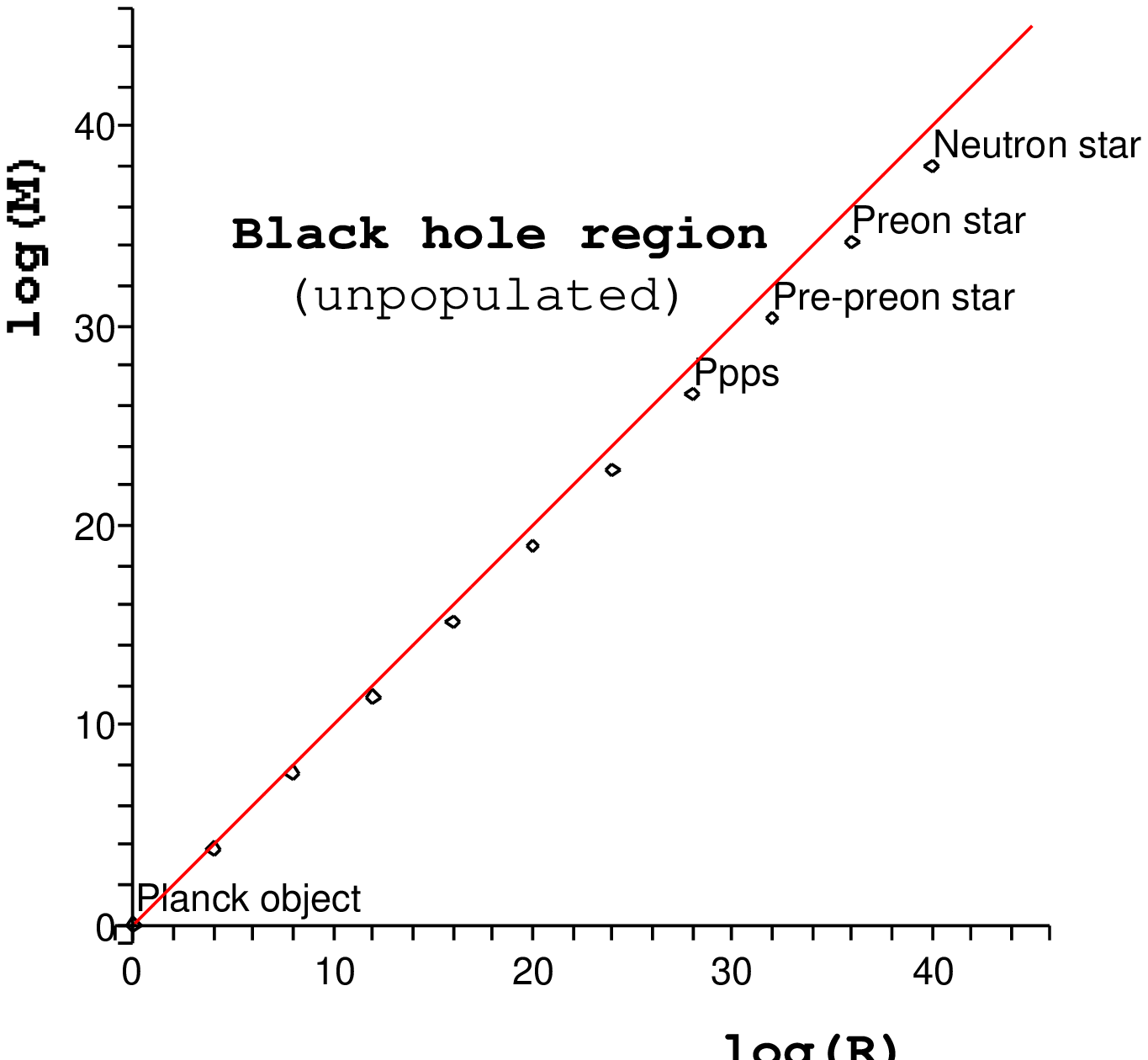}
\end{center}
 \caption{A mass-radius plot of the hierarchy of cosmic compact
 objects. These objects are held up by quantum mechanical
 degeneracy pressure from the "fundamental" constituents (in parenthesis) relevant
 at their scale: Neutron stars
 (neutrons), Preon stars (preons), Pre-preon
 stars (pre-preons), ..., Planck objects (possibly superstrings).
 The stable objects are shown with their maximum mass and minimum size, i.e. for their maximum density, and are separated by regions of unstable configurations (like
 between white dwarfs and neutron stars). The actual data-points
 are thus the end-points of the stable branch in question.
 Also inserted is the region for Schwarzschild, non-rotating, black
 holes. (The region for a rotating, Kerr, black hole is somewhat less
 restrictive.)
Masses and sizes are expressed in dimensionless units as in Fig.1.
The only object bordering on becoming a black hole is the
Planck-object itself.
 According to string theory, and also other candidate theories for quantum gravity, this is related to a
 fundamental minimum
 length in nature, a further increase in energy will \textit{not} resolve smaller scales, due to string duality \cite{Polchinski}.
 The indicated self-similar plot shown will be altered by the exact nature of density fluctuations in the early universe, and the "elementary"
 constituents relevant on that scale, including their detailed interactions. If the normal scenario of inflation remains intact,
 it will dilute all objects originating above the inflation energy scale (e.g. Planck-objects) to unobservable levels in our present universe.}
\end{figure}

\begin{figure}[h]
\begin{center}
\psfig{file=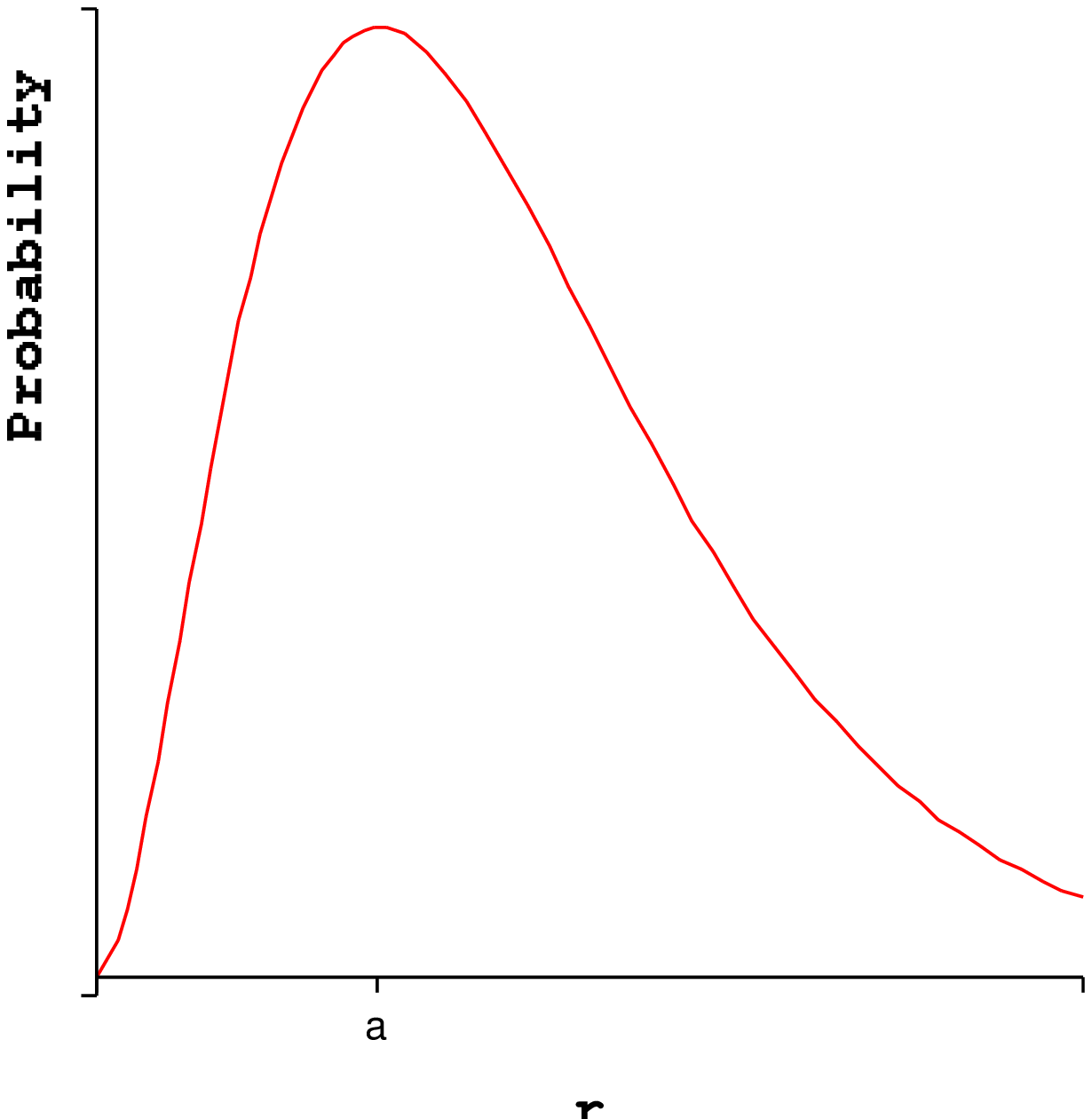}
\end{center}
 \caption{The probability density for the position of the electron in a hydrogen atom. In the ground state (as shown)
it only depends on the radial distance, $r$, from the atomic
nucleus. The height of the curve at a given $r$ gives the
probability that the electron is in that (infinitesimal) interval.
The probability peaks at the Bohr-radius, $a$, the radius of the
innermost circular orbit in the "old" semi-classical Bohr model.
For the position of the event horizon of a black hole an analogous
relation must hold. A problem is that all parts of the
gravitational field in general relativity cannot be represented by
a simple potential, unlike the case for the Coulomb potential in
the hydrogen atom. Neither the Bohr-radius in the hydrogen atom
nor the classical event horizon radius for a black hole are of any
fundamental importance in a truly quantum mechanical treatment.}
\end{figure}

Maybe the most important part of suggesting a novel picture of the
world is to be able to test, and potentially falsify it via
experiments and/or observations.

A first realization is that the observational verification of
\textit{any} compact objects smaller and much more dense than
neutron stars would be a direct "smoking gun" proof of physics
beyond the standard model (especially in the mass-region $M <
10^{12}$ kg excluded for hypothetical, presently existing
primordial black holes \cite{Carr}). Also, such primordial
remnants, together with any sub-constituent "neutrino-like"
radiation, would be potentially observable messengers from the
very early universe, contrasted with today's situation where we
cannot see further back than to a redshift of $z \sim 1100$, or
400 000 years after "time zero" according to the presently favored
cosmological model.

There will be effects in structure formation as i) dark matter is
generated in successive steps with smallest/lightest objects
first, ii) the particle degrees of freedom keep changing, and the
average density of any region in the universe will be
scale-dependent in a hierarchical cosmology as already noted in
\cite{deV}, which changes the expansion rate and dynamics of the
universe, especially at early stages, iii) structure forms from
many more tiny stable\footnote{Our compact objects are unaffected
by the Hawking radiation which evaporates traditionally
conjectured small primordial black holes before they become
effective in structure formation.} "seeds" than traditionally
believed which means more power for small scales in the normal
hierarchical "bottom-up" scenario of structure formation in which
smaller aggregates successively build up larger structures through
gravitational clumping.

Also, there will be quite unique gravitational lensing effects.
The very small and extremely dense objects in the hierarchy will
produce gravitational femto-, pico-, ... lensing events
\cite{HanssonSandin2}. Any such observations would be a
confirmation of our model, while their absence could be used to
refute it.

There would also in principle be deviations from the expected
cosmic microwave background radiation spectrum for very high
multipoles, but in practice it will be undetectable.

Perfect scale-invariance means that density fluctuations will have
equal strength on all scales. In the early universe compact
objects will successively freeze-out when the characteristic
energy, due to expansion of the universe, falls below the relevant
scale of the "elementary" constituents in question. The lightest
compact objects will freeze-out first, and so on, up to Preon
stars which freezes out last. Whatever is left (the final
freeze-out) will not create primordial compact objects, but will
become baryonic matter. If we, just for the sake of argument,
assume the "democratic" principle that the total mass in each
epoch of freeze-out is the same\footnote{It is possible to deduce
a model for primordial density fluctuations where this is
fulfilled exactly: as the abundance of smaller objects naturally
is higher in a fractal-like distribution, coupled with the fact
that the density is larger for earlier epochs, it can be made to
exactly compensate for their lighter masses. However, as such a
fine-tuned model in all probability would not be realized in the
real universe, it is of limited value. An analogous relation also
holds for normal (non-dark) visible matter: the total mass in free
hydrogen (playing the role of our compact "planck-object" for
normal matter) is of the same order of magnitude as that
gravitationally bound in stars in the universe.}, we get the
following relation for the number density, $n_i$, of compact
object class $i$ in the present universe, $n_i = (m_h/m_i) n_h$,
where $m_h$ = mass of hydrogen atom, $m_i$ = typical mass of
compact object class $i$, $n_h$ = mean number density of hydrogen
in present universe $\sim 10^{-7}$/cm$^3$. (For Planck objects,
$m_h/m_{Planck}  \simeq 10^{-19}$, which would be the most
abundant compact object unless the inflationary phase dilutes this
to unobservable levels. For sub-constituent energy scales below
the inflationary scale the formula for $n_i$ can be taken as a
very rough guesstimate of the number density of hierarchy
objects.) In our case, the bulk mass of the universe also departs
more rapidly from thermal equilibrium than in the traditional
model, \textit{one} of three conditions needed to explain the
matter/antimatter asymmetry in the universe \cite{Sakharov_CP}.

As the number of compact object classes due to the scale-invariant
freeze-out in Fig.2 is about ten, we get that dark matter should
be roughly an order of magnitude as abundant as ordinary baryonic
matter, as is observed in astrophysical data. The normal big-bang
constraint on total matter does not apply to the very compact
hierarchy objects, as they have decoupled long before the
expansion (cooling) of the universe and primordial nucleosynthesis
converts gravitationally non-trapped "exotica" into normal
baryonic matter.

To summarize, the main points in this article are:

1) An assumption that there exist sub-constituents on many levels
between the standard model (of particle physics) scale and the
level of grand unified theories and/or superstring theory.

2) The sub-constituents give rise to a hierarchy of stable cosmic
compact objects. Each new sub-constituent level giving rise to
objects lighter, but more dense than the previous scale.

3) Black holes never form. Compact objects on the Planck-scale are
equal to the sub-constituents themselves, terminating the
hierarchy. Here, the compact object for the first time approaches
the border delineating compact objects from black holes.

4) The hierarchy has observational consequences, e.g. in
astrophysical structure formation and gravitational lensing. In
addition, there surely are several potential phenomena and
observational effects which have escaped the author.

\end{document}